\documentclass[useAMS,usenatbib]{mn2e}
\usepackage{graphics,epsfig,aas_macros}
\usepackage{graphicx}
\usepackage[normalem]{ulem}
\usepackage[dvipsnames]{color}
\usepackage[]{inputenc,amssymb}
\usepackage{amsmath}
\usepackage{color}
\usepackage{textcomp}
\usepackage{array}
\usepackage{textcase}
\usepackage{titlesec}
\usepackage{etoolbox}
\usepackage{comment}
\usepackage{caption}
\usepackage[justification=centering]{caption}

\makeatletter
\patchcmd{\ttlh@hang}{\parindent\z@}{\parindent\z@\leavevmode}{}{}
\patchcmd{\ttlh@hang}{\noindent}{}{}{}
\makeatother
\renewcommand\footnoterule{\kern 5pt \hrule width 2in \kern 5.0pt}

\titleformat{\paragraph}
{\it\normalsize\normalsize}{\theparagraph}{1em}{}
\titlespacing*{\paragraph}
{0pt}{3.25ex plus 1ex minus .2ex}{1.5ex plus .2ex}

\newcolumntype{P}[1]{>{\centering\arraybackslash}p{#1}}

\usepackage[title]{appendix}

\def \be{\begin{equation}}
\def \ee{\end{equation}}
\def \bea{\begin{eqnarray}}
\def \eea{\end{eqnarray}}

\title[Precipitation-limit]{Constraints on precipitation-limited hot halos from massive galaxies to galaxy clusters}

\author [P. Singh et al.]
{Priyanka Singh$^{1,2}$ \thanks{priyanka.singh@inaf.it}, G. M. Voit$^3$ and Biman B. Nath$^4$
	\\\\
	$^1$ INAF-Osservatorio Astronomico di Trieste, via G. B. Tiepolo 11, I-34143 Trieste, Italy\\
    $^2$ IFPU - Institute for Fundamental Physics of the Universe, via Beirut 2, 34014 Trieste, Italy\\
    $^3$ Department of Physics \& Astronomy, Michigan State University, East Lansing, MI 48824, USA\\
    $^4$ Raman Research Institute, Sadashiva Nagar, Bangalore 560080, India\\
}

\begin{document}
	\label{firstpage}
	\maketitle
	
	\begin{abstract}
    We present constraints on a simple analytical model for hot diffuse halo gas, derived from a fit spanning two orders of magnitude in halo mass ($M_{500} \sim 10^{12.5}-10^{14.5} M_{\odot}$). The model is motivated by the observed prevalence of a precipitation limit, and its main free parameter is the central ratio of gas cooling timescale to free-fall timescale ($t_{\rm cool}/t_{\rm ff}$). We use integrated X-ray and thermal Sunyaev-Zel’dovich observations of the environments around massive galaxies, galaxy groups and clusters, averaged in halo mass bins, and obtain the best-fitting model parameters. 
    We find $t_{\rm cool}/t_{\rm ff} \sim 50-110$, depending on the model extrapolation beyond the halo virial radius and possibly on biases present in the data-sets used in the fitting analysis.
    The model adequately describes the entire mass range, except for intermediate mass halos ($M_{500} \sim 10^{13.5} M_{\odot}$) which systematically fall below the model predictions. However, the best fits for $t_{\rm cool}/t_{\rm ff}$ substantially exceed the values typically derived from X-ray observations of individual systems ($t_{\rm cool}/t_{\rm ff} \sim 10-30$).  We consider several explanations for those discrepancies, including X-ray selection biases and a potential anti-correlation between X-ray luminosity and the central galaxy's stellar mass.
	\end{abstract}
	
	
\section{Introduction}
Diffuse gas in virialized structures in the universe, such as galaxies or galaxy clusters, offers a powerful probe of the formation and evolution of these structures. While the physical properties of the intracluster medium (ICM) hold clues for the evolution and formation of galaxy clusters, the same can be said about the circumgalactic medium (CGM) vis-a-vis galaxies. The physical processes occurring in the diffuse gas in these two cases are, however, different. In the case of gas in galaxy groups and clusters, the predominant non-gravitational heating sources are likely to be energetic outflows from the central active galactic nuclei (AGN) \citep{boehringer93, mittal11, gaspari12, fabian12}, while  the dominant heating agents of the CGM in galaxies like the Milky Way are thought to be starburst-driven winds \citep{borthakur13,suresh15,suarez16}. In both cases, a feedback loop is needed to keep the hot-gas atmosphere from cooling, condensing to the central part of galaxy and producing too many stars.  Energetic events must heat the gas, and if that gas can cool and fall back into the central galaxy, it should trigger additional energetic events. 

Both the scaling of ICM X-ray luminosity ($L_X$) with halo mass and the effects of AGN feedback on that scaling have been discussed extensively in the literature (e.g., \citealt{churazov01, nath02, brighenti03, dalla04, omma04, Voit2005}). 
The observed relationship between $L_X$ and halo mass implies that feedback in lower-mass halos has a greater impact on the central gas, elevating its cooling time above 
the simple expectations from the radiative cooling of gravitationally heated gas.
Numerous observational studies find that the central cooling time of the ambient medium is tightly correlated with multiphase gas, star formation, and AGN feedback in the central galaxy \citep{Cavagnolo2008, Rafferty2008, Loubser2016, Liu2019, olivares19}, implying that feedback is somehow linked to the central cooling time and the accumulation of multiphase gas. 

Theoretical studies indicate that there should be a lower limit on the central cooling time of the ambient gas, below which it becomes overly susceptible to production of cold gas clouds that rain toward the center and trigger strong feedback \citep{PizzolatoSoker2005,McCourt2012,Sharma2012_tctff, sharma12,gaspari12,Gaspari2013,Gaspari2015,Li2014, Choudhury2015,voit17,Voit2018_turbulence}.  This so-called precipitation limit is near $t_{\rm cool}/t_{\rm ff} \approx 10$, where $t_{\rm cool}$ is the gas cooling timescale and $t_{\rm ff}$ is the free-fall timescale.  Here we adopt the standard definitions in the literature: $t_{\rm cool} \equiv 3nkT / [2 n_e n_i \Lambda(T,Z)]$ and $t_{\rm ff} = (2r/g)^{1/2}$, where $T$ is gas temperature, $n_e$ is electron density, $\Lambda(T,Z)$ is the radiative cooling function for gas of metallicity $Z$, $g$ is the local gravitational acceleration, and $r$ is the distance to the bottom of the potential well. Both observations \citep{Voit2015_Nature,Voit2015_sweeping,voit18,Hogan2017} and simulations \citep{Li2015,Prasad2015,Prasad2018,Meece2017} support the hypothesis that such a limit exists in halos of mass $10^{13}-10^{15} M_{\odot}$, and it may extend all the way down to the mass scale of the Milky Way \citep{Voit2015_precip,voit19,Voit2019_PCGM}.

The limit is thought to arise because accumulation of cold clouds through precipitation dramatically boosts AGN feedback power.  When $t_{\rm cool}/t_{\rm ff}$ in the ICM is far above the limit, there is no multiphase gas fueling the AGN, and feedback is too weak to compensate for radiative cooling. The central gas therefore cools and becomes denser, which lowers $t_{\rm cool}$.  As $t_{\rm cool}/t_{\rm ff}$ approaches the precipitation limit, the ambient medium becomes increasingly susceptible to multiphase condensation, because buoyancy is less able to suppress thermal instability.  When clumps of gas start precipitating out of the ambient medium, they fall toward the center and strongly boost the AGN's fuel supply.  In response, a large increase in AGN power adds heat to the ICM, which eventually raises $t_{\rm cool}$ and alleviates the precipitation.  However, the AGN outburst itself can temporarily stimulate additional condensation and precipitation by lifting low-entropy gas to greater altitudes, producing large perturbations in which $t_{\rm cool}/t_{\rm ff}$ locally drops below unity. The details of this interpretation of the observed floor at $t_{\rm cool}/t_{\rm ff} \approx 10$ are still actively debated \cite[e.g.][]{McNamara2016, Choudhury2019}, but there is general agreement that uplift, non-linear perturbations, and condensation of cold gas clouds are all important components of the feedback process.

Here we explore the prevalence of the precipitation limit by analyzing a large dataset spanning  two orders of magnitude in halo mass: $10^{12.5} M_{\odot}$ to $10^{14.5} M_{\odot}$.  Our approach is inspired by \cite{singh18}, who used joint X-ray and thermal Sunyaev-Zel'dovich (tSZ) effect \citep{sunyaev1972} constraints from stacking of a cosmological sample of massive galaxies \citep{planck13,anderson15} to constrain the hot CGM mass fraction and its temperature. Those two probes are complimentary tracers of the hot gas because they depend differently on density and temperature; X-ray emission depends mainly on the square of the gas-density profile, while the tSZ effect shifts the spectrum of the cosmic microwave background (CMB) through inverse Compton scattering of CMB photons in proportion to the product of density and temperature \citep{sz69}. Together, they constrain both the total amount of ICM/CGM gas and its radial density profile, enabling tests of the precipitation limit on halo mass scales far below those probed by observations of individual objects.  However, \cite{singh18} considered only a limited subset of the overall mass range ($10^{12.6} M_{\odot} \lesssim M_{500} \lesssim 10^{13} M_{\odot}$) because their model, which assumed isothermal gas with a simple power-law density profile ($\propto r^{-1.2}$), was too simple to characterize the hot gas over the full ICM/CGM mass range.

This paper presents a more physically motivated model for the hot halo gas, inspired by the analyses of the precipitation limit, and applies it to the full mass range covered by the joint X-ray and tSZ data sets. The two data sets follow the same selection criteria to calculate the average signal corresponding to a given mass bin from a large statistical sample of galaxies and therefore, are best suited for a combined X-ray-tSZ analysis. Besides, to our knowledge, they are the only available measurements spanning a large dynamical mass range, thus enabling us to test the idea that diffuse gas in both galactic halos and galaxy clusters is maintained in a state dictated by the precipitation limit.

We find that this simple parametric model is able to adequately fit both sets of observational data, thus unifying the ICM and CGM.  However, the best-fitting value of $\min(t_{\rm cool}/t_{\rm ff})$ is 2--3 times greater than indicated by high-quality X-ray observations of the higher-mass objects.  This apparent mismatch likely results from a combination of two different observational biases: (1) individual objects selected from X-ray surveys are likely to have greater central gas density and lower $\min(t_{\rm cool}/t_{\rm ff})$ than is typical for their halo mass because of Malmquist bias, and (2) the mean X-ray luminosity of a stack of central galaxies binned according to their stellar mass is likely to underestimate the X-ray luminosity assigned to a given halo mass because of Eddington bias, which leads to a systematic overestimate of $\min(t_{\rm cool}/t_{\rm ff})$.

The paper is organized as follows. Section \ref{sec-model} presents the parametric model we use in our joint fits to the X-ray and tSZ data sets.  Section \ref{sec-szxr} provides more details about those data sets.  Section \ref{sec-analysis} describes our fitting analysis and presents the results.  Section \ref{sec-discussion} discusses those results, presenting conparisons with other constraints on the ICM and CGM density profiles and exploring potential biases.  Section \ref{sec-summary} summarizes the paper. We adopt WMAP7 best fit cosmological parameters \citep{komatsu11}.

\section{Precipitation-limited hot halo model}
\label{sec-model}

Here we investigate the precipitation-limited model for hot diffuse halo gas introduced by \cite{voit18} and improved in
\cite{voit19}. The initial version of the model assumed that the hot halo gas is in hydrostatic equilibrium within an isothermal potential well having a constant circular velocity.  \cite{voit18} showed that applying an upper limit on electron density, so that $t_{\rm cool}/t_{\rm ff} \geq 10$ at all radii, could fairly reproduce the observed upper limit on the X-ray luminosity-temperature relation across a wide range of halo mass ($10^{12}-10^{15} M_{\odot}$). 

\cite{voit19} made this very simple model more realistic by introducing a decline in the circular velocity profile at large radii, motivated by the observed temperature decline in the outskirts of galaxy clusters  \citep{ghirardini19}.  At large radii, the circular velocity $v_{\rm c}(r)$ of the dark-matter potential well follows an NFW form
\begin{equation}
  v_{\rm c,NFW}(r) =
      v_{\rm c, max} \sqrt{4.625 \left[ \frac{\ln(1+r/r_s)}{r/r_s} - \frac{1}{1+r/r_s}\right]} 
\label{eqn-vcNFW}
\end{equation}
where $v_{\rm c,max}$ is the potential's maximum circular velocity, the factor 4.625 ensures that $v_{\rm c}(r_s) = v_{\rm c, max}$ for an isothermal potential  and $r_s$ is a scale radius corresponding to the concentration parameter $c(M,z) = R_{200} / r_s$\footnote{$M_{\Delta} \equiv \Delta \times 4\pi \rho_c R^3 _{\Delta}/3$, where $\rho_c$ is the critical density of the Universe.}.  
\cite{voit19} assumed $v_{\rm c}$ to be constant at small radii, in order to approximate a galactic potential well.  Those functional forms were then continuously joined at the radius $r_{\rm max} = 2.163 r_s$ (independent of the concentration parameter), where the circular velocity of an NFW profile peaks, giving 
\begin{equation}
  v_{\rm c}(r) =
    \begin{cases}
      v_{\rm c, NFW} & , \; \; \text{$r > r_{\rm max}$}  \\
      v_{\rm c, max} & , \; \; \text{$r < r_{\rm max}$}  \; \; .
    \end{cases}       
\label{eqn-vc}
\end{equation}
However, this form is inaccurate for galaxy clusters, because $v_{\rm c,max}$ can greatly exceed the circular velocity of the central galaxy, which is usually not much greater than $\sim 400 \, {\rm km \, s^{-1}}$.  In this paper, we therefore use a different potential model for the halos with $v_{\rm c,max} > 400 \, {\rm km \, s^{-1}}$.  For these high-mass halos, we limit the constant portion of the inner circular-velocity profile to be no greater than 400 km s$^{-1}$, 
so that
\begin{equation}
  v_{\rm c}(r) =
    \begin{cases}
      v_{\rm c,NFW}(r) & , \; \; r > r_{\rm max} \\
      \max [ v_{\rm c,NFW}(r) , 400 \, {\rm km \, s^{-1}}]  
        & , \; \; r < r_{\rm max} \; \; .
    \end{cases}       
\label{eqn-vc_400}
\end{equation}
In all of our models, the concentration parameter depends on halo mass according to \cite{dutton14}.

The properties of the ambient hot halo gas in a spherical potential well are fully determined by its radial profile of specific entropy, the assumption of hydrostatic equilibrium, and a boundary condition.  In the literature on this subject, specific entropy is typically represented in terms of the entropy index $K \equiv kTn_e^{-2/3}$. We assume here that the entropy profile follows the precipitation-limited NFW (pNFW) model of \cite{voit19}: 
\begin{equation}
  K_{\rm pNFW}(r) = K_{\rm base}(r) + K_{\rm pre}(r)
  \; \; ,
\end{equation}
where
\begin{equation}
  K_{\rm base}(r) = 1.32 \, \frac{k T_{\phi}(r_{200})}{\bar{n}^{2/3}_{e,200}} 
                        \left( \frac{r}{r_{200}}\right)^{1.1} \; \; ,
\end{equation}
\begin{equation}
  K_{\rm pre}(r) = (2 \mu m_p)^{1/3} \left[\frac{t_{\rm cool}}{t_{\rm ff}} \frac{2n_i}{n}
        \frac {\Lambda(2T_{\phi})} {3} \right]^{2/3} r^{2/3}
        \; \; .
\end{equation}
In these expressions, $k T_{\phi} \equiv \mu m_p v^2_c(r)/2$ is the gravitational temperature associated with $v_c$ and $\bar{n}_{e,\rm 200}$ is the mean electron density associated with a total matter overdensity 200 times the critical density.  Those characteristic values of temperature and density set the scale of the entropy profile $K_{\rm base}(r)$ produced by cosmological structure formation \citep[e.g.,][]{Voit2005}.  The precipitation-limited entropy profile $K_{\rm pre}(r)$ is determined by choosing a value for the ratio $t_{\rm cool}/t_{\rm ff}$.  In this paper, we allow that ratio to be a free parameter, determined by fitting the data. The definition of $K_{\rm pre}$ assumes $T = 2 T_{\phi}$ because that is the hydrostatic temperature of the gas with $K \propto r^{2/3}$ in an isothermal potential.

Integration of the equation of hydrostatic equilibrium, given $K(r)$ and $v_c(r)$, yields 
\begin{eqnarray}
T_{\rm pNFW}(r) & = & \left[ \frac {K_{\rm pNFW}^{3/5}(r)} {K_{\rm pNFW}^{3/5}(r_{200})}  \right] T(r_{200}) 
   \nonumber \\ 
    &  & + \, \frac{4}{5} \int^{r_{200}}_r 
                   \left[ \frac {K_{\rm pNFW}^{3/5}(r)} {K_{\rm pNFW}^{3/5}(r^\prime)} \right] T_{\phi}(r^\prime) 
                   \, \frac {d r^\prime} {r^\prime} 
    \; \; .
    \label{eq-kTpNFW}
\end{eqnarray}

\noindent Obtaining a particular solution therefore requires a boundary condition, which we choose to be the gas temperature at $r_{200}$, specified by 
\begin{equation}
    T(r_{200}) = f_T 
            \left( \frac{M_{200}}{10^{14} M_{\odot}} \right)^{\alpha_T} 
            \times T_{\phi}(r_{\rm 200})
\end{equation}

\noindent This expression for the temperature boundary condition introduces two more free parameters into the overall model:  $f_T$ is the ratio of the temperature at $r_{200}$ to the virial temperature, $T_\phi(r_{200})$, of a halo of mass $M_{200} = 10^{14} \, M_\odot$, and $\alpha_T$ determines the dependence of that ratio on halo mass.

The only degree of freedom remaining in the model is the halo gas metallicity, which is implicit in the radiative cooling function ($\Lambda$) used to define $K_{\rm pre}$.  There is a large scatter in the observed metallicity of the ICM \citep{leccardi08, molendi16, mernier17, lovisari19}.  However, all of these observations point towards a rather flat metallicity profile beyond $0.3-0.4 R_{500}$, with a rise in metallicity near the cluster centre. Determining  the metallicity of hot CGM gas in lower-mass halos is even more difficult due to the lower gas densities. \cite{prochaska17} found a median metallicity for the cool CGM of low redshift $L^*$ galaxies close to $0.3Z_{\odot}$. \cite{Oppenheimer2020} found the metallicity of hot CGM gas in the EAGLE simulation to be greater than $0.1 Z_{\odot}$.
Given the large uncertainties in halo gas metallicity and the relative insensitivity\footnote{We verify that our results remain largely unaffected in the gas metallicity range $\sim 0.1-1Z_{\odot}$.} of tSZ and $L_{0.5-2 \, {\rm keV}}$ on gas metallicity, we fix $Z=0.3$, consistent within the observational uncertainty in a wide radial range.  With this assumption, Equation \ref{eq-kTpNFW} gives a unique temperature profile, from which we obtain radial profiles of electron density $n_e(r) = [ kT_{\rm pNFW}(r) / K_{\rm pNFW}(r) ]^{3/2}$ and electron pressure $P_e(r) = n_e(r) k T_{\rm pNFW}$.

To summarize, the free parameters of the model are, (i) $t_{\rm cool}/t_{\rm ff}$, the ratio of gas cooling timescale to free-fall timescale, (ii) $f_T$, the ratio of gas temperature at $r_{200}$ to the virial temperature at $M_{200} = 10^{14} \, M_{\odot}$, and (iii) $\alpha_T$, the mass-slope of the ratio $T(r_{200})/T_{\phi}(r_{\rm 200})$.

\section{X-ray and tSZ data sets}
\label{sec-szxr}

The tSZ signal traces gas pressure integrated along the line of sight, and hence is degenerate between gas density and temperature.  It is most useful for measuring the product of halo gas mass and  temperature.  The X-ray signal depends primarily on the integral of gas density squared along the line of sight and has a weaker dependence on gas temperature.  It therefore helps to break the tSZ degeneracy between gas density and temperature and is sensitive to the distribution of gas within the halo, particularly at small radii, where gas density peaks.  Combination of the two signals has been previously used to investigate cluster scaling relations \citep{planck11, dietrich19} and the detailed gas distribution and thermodynamics in individual systems \citep{sayers18, shitanishi18} as well as averaged profiles using stacked measurements \citep{singh15b, singh18} and to constrain cosmological parameters (see \citealt{kozmanyan19} and references therein).

The analysis in this paper uses stacked tSZ \citep{planck13} and X-ray measurements \citep{anderson15} which combine a large number of locally brightest galaxies (LBGs) from the New York University Value Added Galaxy Catalogue based on SDSS-DR7, stacked in stellar mass bins to obtain a high signal-to-noise ratio (SNR) from galaxy clusters down to massive galaxies.  Conversion of stellar mass to halo mass is done using a mock SDSS catalog created from the Millennium Simulation \citep{ springel05b, guo13}.
Both the above mentioned tSZ and X-ray studies follow the same LBG selection criterion. The two data-sets uniquely span two orders of magnitude in halo mass, stacking $\sim 22,000$ objects in the lowest mass bin considered in the paper to around 36 objects in the highest mass bin. Stacked measurements from such a large number of objects help in averaging over one-to-one variation among individual objects, thus providing robust estimates of gas properties as a function of halo mass.

\cite{anderson15} stacked X-ray luminosity in the 0.5-2 keV band of the ROSAT all sky survey coming from within $R_{500}$ around the LBG. \cite{planck13} stacked cylindrical Compton $y$-parameter measurements integrated over an aperture of radius $5 \times R_{500}$.  Analytically, these two observable quantities are given by
\begin{equation}
L_{0.5-2 \, {\rm keV}} = \int_{0} ^{R_{500}}
2\pi r dr \int_r ^{R_{500}} \frac{2\, n_e n_i \Lambda(Z,T_e) r'
	dr'}{\sqrt{r'^2-r^2}}
\end{equation}
\begin{equation}
Y_{\rm cyl}=\frac{\sigma_T}{m_e c^2 D^2 _A(z)}  
\int_0 ^{5 R_{500}} 2\pi r dr \int_r ^{5 R_{500}} \frac{2 P_e(r')r'
	dr'}{\sqrt{r'^2-r^2}}
\label{eqn-ycyl}
\end{equation}
Note that \cite{planck13} derive $Y_{R500}$ (i.e. the tSZ signal integrated within $R_{500}$) using the pressure profile from \cite{arnaud10} to convert the observed $Y_{\rm cyl}$ to $Y_{R500}$. The conversion factor is $\sim 1.796$. \cite{brun15} showed that the conversion factor is sensitive to assumptions about the pressure profile. We, therefore, directly compare our estimates to $Y_{\rm cyl}$ to avoid any biases introduced due to the conversion.

We use the Astrophysical Plasma Emission Code (APEC; \citealt{smith01}) to compute the cooling function $\Lambda(T,Z)$ corresponding to the soft X-ray band, 0.5-2 keV.
The tSZ signal is scaled to a fixed angular diameter distance and 
both the tSZ and X-ray signals are then scaled to zero redshift, to give
\begin{equation}
    \tilde{L}_{0.5-2 \, {\rm keV}} = L_{0.5-2 \, {\rm keV}} E^{-7/3}(z)
\end{equation}
\begin{equation}
    \tilde{Y}_{\rm cyl} = Y_{\rm cyl} E^{-2/3}(z) (D_{\rm A}(z)/500 \rm Mpc)^2
    \; \; ,
\end{equation}
where, $D_{\rm A}(z)$ is the angular diameter distance at redshift $z$ and $E(z) = \sqrt{\Omega_{\Lambda}+\Omega_{\rm m}(1+z)^3}$. Computation of $Y_{\rm cyl}$ requires extrapolation of the pNFW model beyond $r_{200}$, where the assumption of hydrostatic equilibrium is less justifiable.  Formally, the temperature boundary condition applied at $r_{200}$ in the pNFW model results in a rapid drop in gas pressure beyond $r_{200}$, which unphysically suppresses the resultant tSZ signal.  We therefore choose to extend the pNFW model beyond $r_{200}$ using two different assumptions that we expect to bracket the actual conditions:
\begin{itemize}
  \item \textbf{$\gamma$--model}. This model for gas beyond $r_{200}$ assumes a power-law pressure profile,  $P_{\rm pl}(r) \propto  r^{-3.3}$ at $r>r_{200}$, which corresponds to $n_e \propto r^{-3}$ with a polytropic index $\gamma = 1.1$. The assumed slope of the gas density profile corresponds to the NFW dark matter profile.
  The adopted value of $\gamma$ is consistent with the range suggested by \cite{komatsu01}, based on requiring the slope of the gas density profile to match the slope of dark matter density profile near the virial radius, and is also in agreement with the observed pressure profile near galaxy cluster outskirts \citep{Finoguenov2001, Atrio2008, McDonald2014}. \\
  
  \item \textbf{Flat~$P$ model.} This alternative model assumes that the gas beyond $r_{200}$ has constant pressure.  That assumption might be more appropriate for lower-mass systems in which feedback has pushed out much of the halo gas that would have been within $r_{200}$, producing a flatter pressure profile at large radii.  
\end{itemize}

Together, these models are likely to bracket the actual pressure profiles beyond $r_{200}$, and we normalize both of them so that the pressure at $r_{200}$ is continuous with the pNFW profile. Section \ref{sec-fpp} discusses the sensitivity of our best fits to these assumptions about gas pressure at large radii.

The halo mass estimates for both \cite{planck13} and \cite{anderson15} depend on modelling uncertainties in the galaxy population because their relationships to each stellar mass bin were determined through forward modelling of sample selection and signal measurement using a single galaxy-population simulation from \cite{guo13}. \cite{wang16} later revised those halo mass estimates using stacked weak gravitational lensing measurements, because the lensing signal is more robust to modelling uncertainties. Also, \cite{wang16} accounted for observational and modelling uncertainties by comparing results from many galaxy-population simulations. Their re-calibration resulted in 30\% increases in the amplitude of the tSZ scaling relation and a 40\% increase in case of the X-ray scaling relation. Therefore, we use the effective halo masses from \cite{wang16} in our analysis.

\section{Fitting analysis and results}
\label{sec-analysis}

We used MCMC analysis to fit the pNFW models to the tSZ and X-ray data sets. The free parameters, $t_{\rm cool}/t_{\rm ff}$, $f_T$ and $\alpha_T$, are described in Section \ref{sec-model}. The total log-likelihood function for the combination of these two data sets (which we assume are independent and uncorrelated at fixed halo mass) is given by, 
\begin{equation}
\begin{split}
  \ln \mathcal{L} = C 
        - \frac{1}{2} \left[ \sum_{i=1}^{N_{\rm sz}} 
                \left( \frac{\log \tilde{Y}_{\rm cyl} - \log \tilde{Y}^{{\rm obs}}_{\rm cyl}}
                            {\sigma_{\log \tilde{Y}_{\rm cyl}^{\rm obs}}}
                            \right)_i^2  \right.
    \\  
        + \left. \sum_{j=1}^{N_{L_{\rm x}}} 
                 \left( \frac{\log \tilde{L}_X - \log \tilde{L}^{{\rm obs}} _X}
                             {\sigma_{\log \tilde{L}_X^{\rm obs}}} 
                             \right)_j^2 \right]
\label{eqn-lnlike}
\end{split}
\end{equation}
where the summations ($i$ and $j$) are over the halo mass bins ($M^{i/j} _{500}$) in which the tSZ/X-ray measurements are stacked, and $\sigma_{\log \tilde{Y}_{\rm cyl}^{\rm obs}}$ and $\sigma_{\log \tilde{L}_X^{\rm obs}}$ are the observed log-normal uncertainties in the tSZ-mass and luminosity-mass scaling relations, respectively. The log-normal uncertainty $\sigma_{\log O}$ (where $O \equiv \tilde{Y}_{\rm cyl}^{\rm obs}$ or $\tilde{L}_X^{\rm obs}$) includes contributions from uncertainty in both the observable $O$ (defined as $\sigma _{\log O, \rm signal}$) and the determination of $M_{500}$ (defined as $\sigma_{\log M_{500}}$). It is given by,
\begin{equation}
    \sigma_{\log O} = \sqrt{\sigma^2 _{\log O, \rm signal} + \sigma^2 _{\log M_{500}} \Bigl( \frac{d \log O}{d \log M_{500}}\Bigr)^2}.
\label{eqn-sigmas}
\end{equation}
For $\sigma _{\log O, \rm signal}$, we use the uncertainties calculated using bootstrap analysis (see Table 1 and Table 3 in \citealt{planck13} and \citealt{anderson15}, respectively). We refer readers to \cite{wang16} for the details of uncertainty estimation in case of $M_{500}$. Note that the number of halo mass bins used while fitting the tSZ and X-ray data sets are different i.e. $N_{\rm sz} \neq N_{\rm Lx}$ due to the low signal-to-noise in tSZ for the lowest halo mass bin. The term 
$C= -\frac{1}{2} \left[ \sum \ln (2\pi \sigma^2 _{\log \tilde{Y}_{\rm cyl}^{\rm obs}})+\sum \ln (2\pi \sigma^2 _{\log \tilde{L}_X^{\rm obs}}) \right]$ 
is independent of input model parameters and therefore does not affect the fitting analysis.

It is difficult to observe the temperature profile in low mass systems due to the faint X-ray emission, however, a temperature decline has been observed in a few massive spiral galaxies \citep{anderson16, bogdan17}. Observations of the ICM temperature profile also show a decrease in temperature near cluster outskirts \citep{moretti12, walker12, ghirardini18, ghirardini19}.
Therefore, we assume a conservative Gaussian prior on $f_T = 0.5 \pm 0.25$ (in addition to $f_T>0$). The fiducial value of $f_T$ is motivated by the observed temperature decline, to approximately half the virial temperature near $R_{200}$ in galaxy clusters \citep{ghirardini19}. We apply uninformative uniform priors on $t_{\rm cool}/t_{\rm ff}$ and $\alpha_T$.

\subsection{Best-Fitting Parameters}

\begin{figure}
\hspace*{-0.5cm}
	\includegraphics[height=9cm,angle=0.0 ]{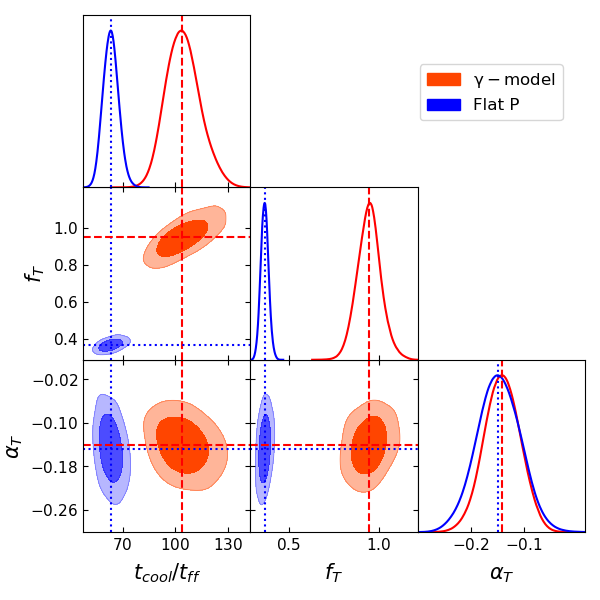}
	\caption{The 68\% and 95\% confidence limit contours for our fiducial $\gamma-$model (red contours) and Flat~$P$ model (blue contours) shown by the dark and light shaded regions, respectively. Dashed-red and dotted-blue lines represent best-fitting parameter values for the $\gamma-$model and Flat~$P$ model, respectively.}
	\label{fig-mcmc}
\end{figure}

Figure \ref{fig-mcmc} shows the resultant one-dimensional and two-dimensional posterior probability distribution
for the $\gamma-$model (red contours) and the Flat~$P$ model (blue contours), with dashed-red and dotted-blue lines highlighting the best-fitting values of the model parameters, respectively. Table \ref{tab-ss} gives the best-fitting values of those parameters for both the $\gamma$--model and the Flat~$P$ model, along with their values when the data points near halo mass $\sim 10^{13.5} \, M_\odot$ are excluded.

For the $\gamma$--model, the best-fitting values of the model parameters are $t_{\rm cool}/t_{\rm ff} = 104.06^{+9.74}_{-9.40}$, $f_T=0.95\pm0.06$ and $\alpha_T = -0.14\pm0.03$. For the Flat~$P$ model, the best-fitting values are $t_{\rm cool}/t_{\rm ff} = 63.23^{+4.43}_{-3.34}$, $f_T=0.36\pm0.02$ and $\alpha_T = -0.15\pm0.04$. 
The normalization of the density profile and hence $L_X$ are sensitive to $t_{\rm cool}/t_{\rm ff}$, therefore, the constraints on the value of $t_{\rm cool}/t_{\rm ff}$ are largely determined by the X-ray measurements for a given model. On the other hand, both X-ray and SZ measurements play a crucial role in breaking density-temperature degeneracy and thus constraining $f_T$ as well as $\alpha_T$.
The statistical uncertainties on the parameters $t_{\rm cool}/t_{\rm ff}$ and $f_T$ are formally $<10\%$ for both models, but the model-dependent disagreements are much greater, indicating that the systematic uncertainties in our modeling are greater than the statistical ones. In contrast, however, the best-fitting values of $\alpha_T$ are in good agreement.  Also, $\alpha_T$ is significantly ($ \gtrsim 3\sigma$) smaller than zero, indicating that the ratio of gas temperature near the virial radius to the halo virial temperature declines as the halo mass increases.

\begin{figure}
	\includegraphics[height=13.cm,angle=0.0,trim = 0.0in 0.5in 0.0in 1.0in,clip]{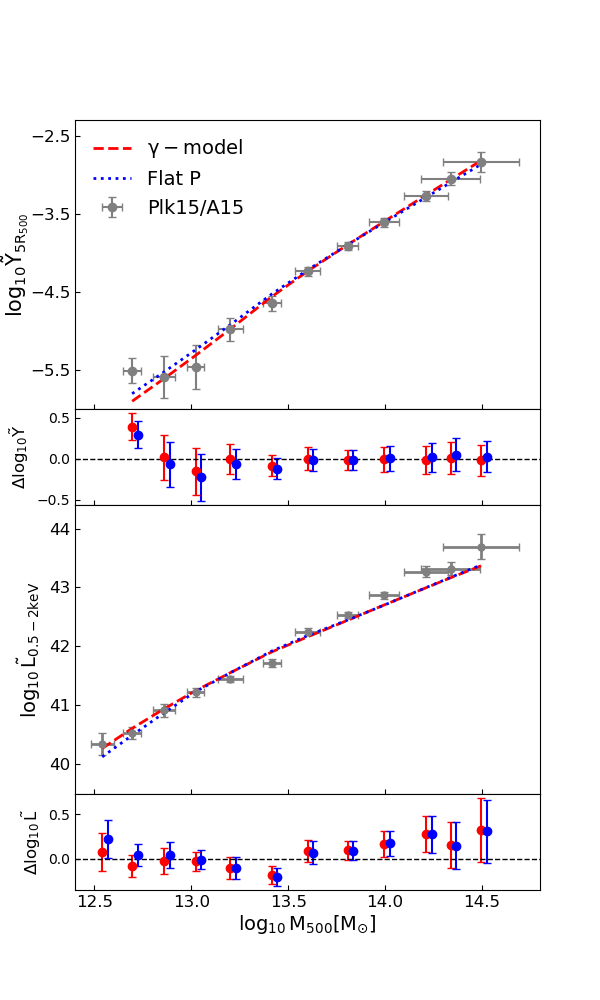}
	\caption{Comparison of best-fitting $\gamma$-model predictions (dashed-red lines) and Flat~$P$ model (dotted-blue lines) with stacked total X-ray luminosity (lower panel) and stacked tSZ (upper panel) observations (grey points). The sub-panels at the bottom of each panel shows the residuals for the $\gamma$-model (red points) and Flat~$P$ model (blue points). The error-bars shown in the residual panels are given by Equation \ref{eqn-sigmas}. The blue points are shifted slightly towards right-hand side for clarity.}
	\label{fig-xr}
\end{figure}

Figure \ref{fig-xr} compares the X-ray and tSZ data sets used in the fitting analysis with the best-fitting model predictions for the $\gamma-$model (dashed-red lines in the main panels and red points in the residual panels) and the Flat~$P$ model (dotted-blue lines in the main panels and blue points in the residual panels). Both models provide  reasonable fits to the entire mass range, capturing the overall trend quite well considering the simplicity of the modeling. The models under-predict the tSZ signal in the lowest mass bin, however, the detection in this bin is at a significance $< 3\sigma$ and the residuals are consistent with zero within 2$\sigma$.

The above analysis assumes that the stacked X-ray/tSZ signals represent individual halos (i.e. one-halo terms only). However, \cite{vikram17} and \cite{hill17} showed that the two-halo term makes a non-negligible contribution to the {\it Planck} measurement of the stacked tSZ signal at masses $M_* < 10^{11.4} M_{\odot}$ i.e. $M_{500} < 10^{13.6} M_{\odot}$. Repeating the analysis without using tSZ data for these low mass bins gives, $t_c/t_{ff} \approx 93 \pm 11$, $f_t \approx 0.89 \pm 0.08$ and $\alpha_T \approx -0.07 \pm 0.06$. The best-fitting values of $t_c/t_{ff}$ and $f_T$ are consistent with the fiducial scenario within 1-$\sigma$ uncertainty, whereas, $\alpha_T$ decreases by a factor of two and is poorly constrained, showing that the low mass bins are crucial in constraining the relationship between gas temperature and halo mass. However, the one-halo terms are expected to be dominant even at lower masses, and therefore our results are robust against the two-halo contamination.
 
The best-fitting models systematically underpredict X-ray luminosity at cluster scales and overpredict it near $M_{500} \sim 10^{13.5} \, M_\odot$. Additionally, the best-fitting values of $t_{\rm cool}/t_{\rm ff}$ are significantly larger than the ones indicated by X-ray selected cluster samples.  Section \ref{sec-discussion} discusses in detail the discrepancy at intermediate masses, the comparison of our best fits with observations of gas-density profiles in individual galaxies and clusters, and the impacts of biases on our results.

\begin{table*}
\caption{Constraints on pNFW model parameters from MCMC analysis. Values in parentheses are constraints obtained when excluding the mass bins $M_{500} = 10^{13.20}$ and $10^{13.42} M_{\odot}$.}
\resizebox{0.72 \textwidth}{!}{
\setlength{\tabcolsep}{5pt}
	\begin{tabular}{c c c c}
			\\
			& $\gamma-$model & $\gamma-$model & Flat $P$ model \\
			& ($\rm P_{pl} \propto  r^{-3.3}$) & ($\rm P_{pl} \propto  r^{-2.2}$) & \\\\
			\hline \\
			$t_c/t_{ff}$ & $104.06^{+9.74}_{-9.40}$ ($91.55^{+10.36}_{-8.38}$) & $89.18^{+7.88}_{-7.22}$ ($79.21^{+7.80}_{-7.40}$) & $63.23^{+4.43}_{-4.34}$ ($56.68^{+4.70}_{-4.23}$) \\\\
			$f_T$ & $0.95\pm0.06$ ($0.90\pm0.07$) & $0.74\pm0.05$ ($0.71\pm0.05$) & $0.36\pm0.02$ ($0.35\pm0.02$) \\\\
			$\alpha_T$ & $-0.14\pm0.03$ ($-0.14\pm0.04$) & $-0.15\pm0.04$ ($-0.14\pm0.04$) & $-0.15\pm0.04$ ($-0.14\pm0.04$) \\\\
			reduced-$\chi^2$ & 0.84 (0.64) & 0.83 (0.93) & 0.50 (0.48) \\\\
			\hline
		\end{tabular}
		\label{tab-ss}}
\end{table*}

\begin{figure*}
\hspace*{-1.0cm}
	\includegraphics[height=12.2cm,angle=0.0]{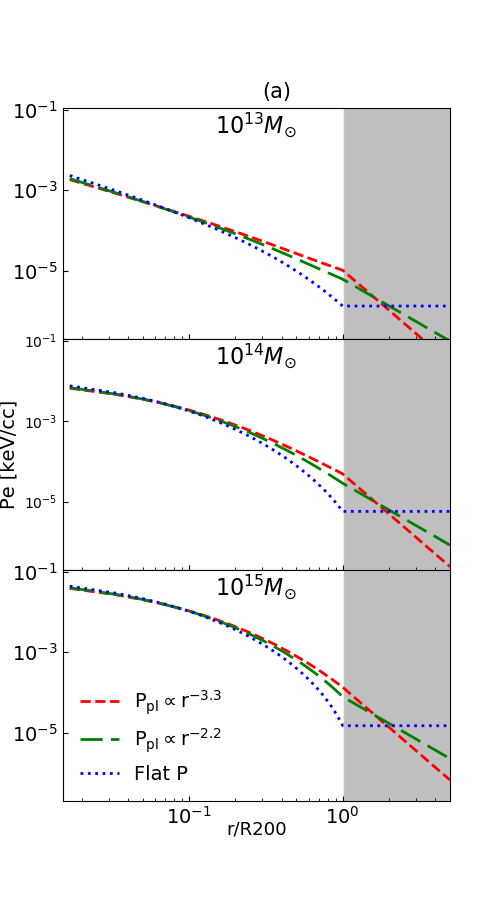}
	\includegraphics[height=12.2cm,angle=0.0]{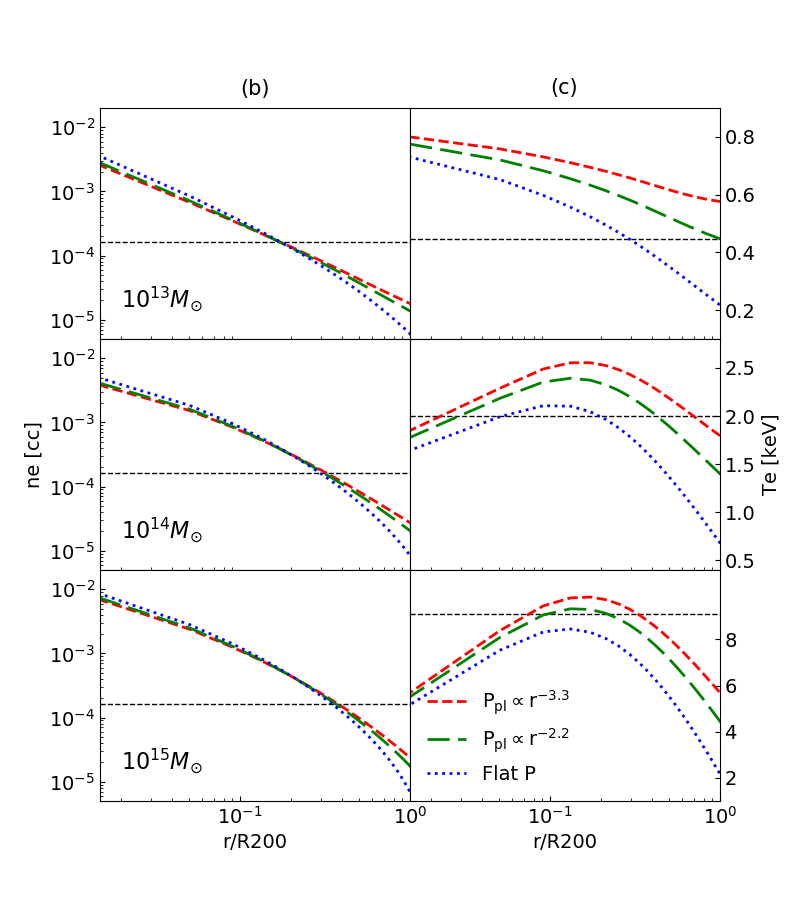}
	\caption{Panel (a): Pressure profiles for the best-fitting $\gamma$-models (dashed-red lines for $P_{\rm pl} \propto  r^{-3.3}$ and long dashed-green lines for $P_{\rm pl} \propto  r^{-2.2}$) and the Flat $P$ model (dotted-blue lines) for three halo masses $M_{500} = 10^{13} M_{\odot}$ (top panel), $10^{14} M_{\odot}$ (middle panel) and $10^{15} M_{\odot}$ (bottom panel). The grey shaded regions highlight the radial range where the pressure profiles are extrapolated. Panel (b) and Panel (c): the density and temperature profiles, respectively, for the best-fitting models. Horizontal dashed-black lines correspond to $\bar{n}_{e,\rm 200}$ in Panel (b) and $T_{\phi}(r_{\rm 200})$ in Panel (c).}
	\label{fig-te}
\end{figure*}

\subsection{Sensitivity to the pressure profile beyond $R_{200}$}
\label{sec-fpp}

Our modeling depends systematically on assumptions about gas pressure beyond $R_{200}$ because of how those assumptions determine the profiles of gas temperature and density at smaller radii. Panel (a) of Figure \ref{fig-te} shows  pressure profiles corresponding to the best-fitting parameters in Table \ref{tab-ss}.  Beyond $R_{200}$, those pressure profiles are constrained entirely by the tSZ observations, which reflect the product of temperature and total gas mass within the tSZ aperture $5 R_{500} (\approx 3 R_{200})$. The Flat~$P$ model has a greater total gas mass within that aperture, and so gas temperature at $R_{200}$ must be lower than in the $\gamma$--model, in order to satisfy the tSZ constraints (i.e. $Y_{\rm cyl}$). Table \ref{tab-ss} shows that the resulting difference in best-fitting temperature at $R_{200}$ (as represented by $f_T$) is approximately a factor of 2.

In our models, this change in the boundary temperature requires a compensating change ($\sim 30\%$) in the $t_{\rm cool}/t_{\rm ff}$ parameter in order to satisfy the X-ray constraints. Panel (b) of Figure \ref{fig-te} shows that $L_X$ constraints force the density profiles of our two best-fitting models to be nearly identical at $\lesssim 0.5 R_{200}$.  However, panel (c) shows that gas temperature in the Flat~$P$ model is systematically lower at all radii. Lower temperature everywhere is a direct consequence of the lower-temperature boundary condition at $R_{200}$, because the entropy profile and gravitational potential are fixed (see equation \ref{eq-kTpNFW}). 

In addition to our fiducial $\gamma$--model (i.e. $P_{\rm pl} \propto  r^{-3.3}$) and Flat~$P$ model, Table \ref{tab-ss} and Figure \ref{fig-te} also present the best-fitting results for an intermediate outer pressure profile slope, namely $P_{\rm pl} \propto r^{-2.2}$ which corresponds to $n_e \propto r^{-2}$. In this case, the best-fitting values of $t_{\rm cool}/t_{\rm ff}$ $(\approx 89.18^{+7.88}_{-7.22})$ and $f_T$ $(\approx 0.74\pm0.05)$ are intermediate between the two bracketing models. Note that, the best-fitting value of $f_T$  for the shallower slope corresponds to the gas temperature closer to its observed value in galaxy cluster \citep{ghirardini18}.

Figure \ref{fig-te} highlights that the density profiles for the two $\gamma-$models are nearly identical within $R_{200}$. The X-ray emission is sensitive to the central density profile (since X-ray emissivity is $\propto n^2 _e$), and therefore the model strongly constrains the shape of the density profile. Thus, the density profiles of best-fit models are all similar, but the best-fit $t_{\rm cool}/t_{\rm ff}$ ratios differ due to the different gas temperatures which is constrained by the tSZ signal. 
Models with steeper pressure profiles add very little to the tSZ signal that comes from within $R_{200}$, because the integral of pressure over volume rapidly converges with increasing radii. Therefore, the temperature of the gas within $R_{200}$ needs to be large enough for that region to supply essentially all the SZ signal.
The two features give us a degeneracy between $t_{\rm cool}/t_{\rm ff}$ and $f_T$ that depends on the outer pressure profile slope.

We expect the actual thermodynamic profiles of the ICM and CGM to be bracketed by our $\gamma$--model and Flat~$P$ model.  Direct observations of the temperature profiles of galaxy clusters suggest that they more closely resemble the $\gamma$--model.  However, the apparent deficiency of baryonic gas mass, relative to the cosmological baryon fraction, in lower-mass halos suggests that the Flat~$P$ model might be more applicable to them, because of its greater fraction of baryonic mass beyond $R_{200}$.

\section{Discussion}
\label{sec-discussion}
\subsection{Comparison with individual systems}
\label{sec-indiv}
\begin{figure*}
\hspace*{-1.0cm}
	\includegraphics[angle=0.0]{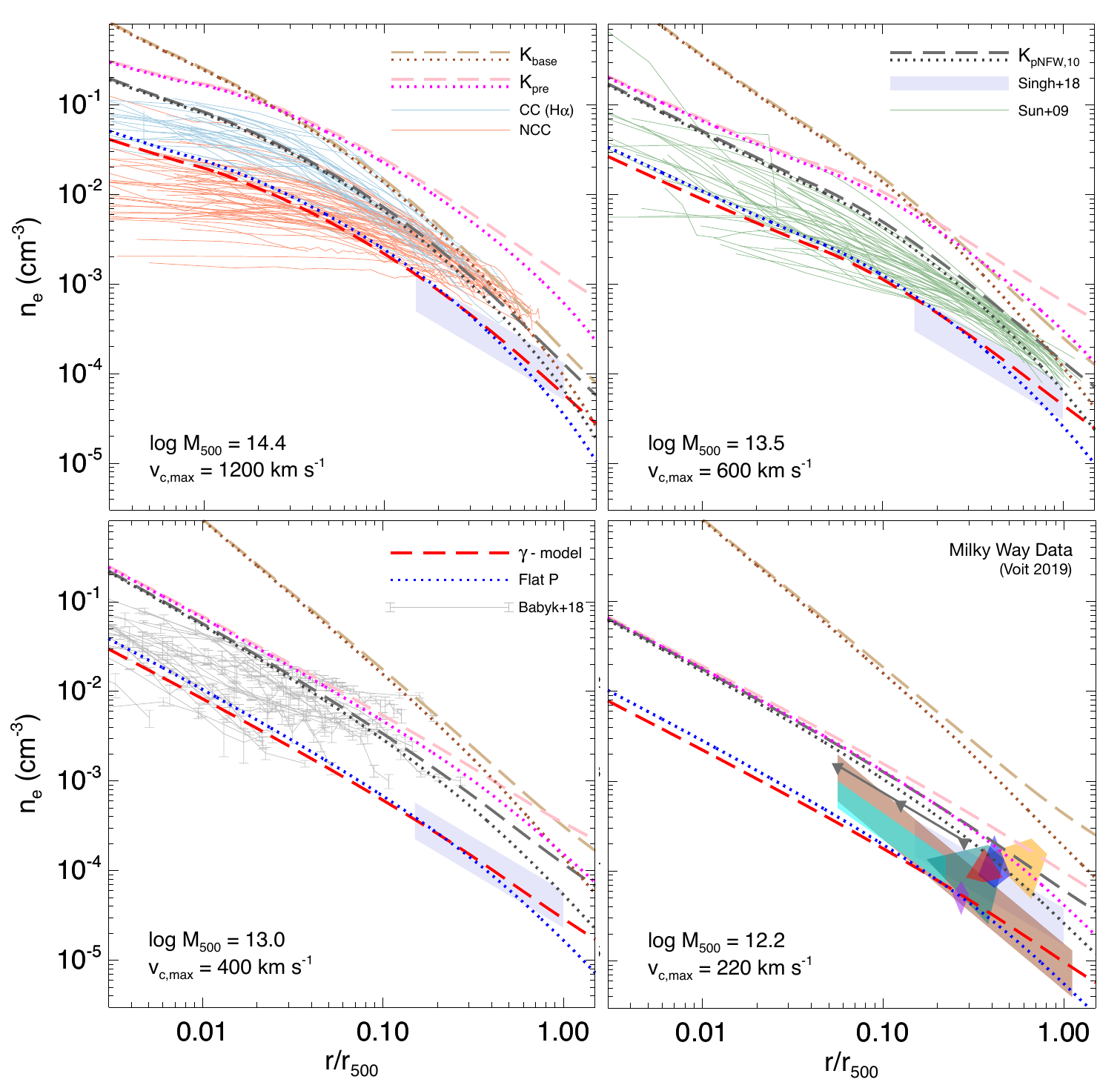}
	\caption{Comparisons of models with observed electron density profiles of individual halos. Top left: Galaxy clusters from the ACCEPT sample (\citealt{cavagnolo09}), with solid blue lines representing cool-core (CC) clusters with central H$\alpha$ emission and solid red lines representing non-cool-core (NCC) clusters without H$\alpha$ emission.  Top right: Galaxy groups from Sun et al. (2009), shown with solid green lines.  Bottom left: Early-type galaxies from Babyk et al. (2018, shown with solid grey lines. Bottom right: Constraints on the Milky Way's ambient CGM, compiled by Voit (2019)  and represented with polygons and a line with inverted triangles (see text for details).  In each panel, thick brown lines show cosmological baseline density profiles (derived from $K_{\rm base}$), thick pink lines show precipitation-limited profiles (derived from $K_{\rm pre}$ with $t_{\rm cool}/t_{\rm ff} = 10$), and thick grey lines show the pNFW profiles obtained from combining $K_{\rm base}$ and $K_{\rm pre}$.  Dashed lines represent profiles with the $\gamma$-model pressure profile at large radii.  Dotted lines represent profiles with the Flat $P$ profile at large radii.  As in Figure~\ref{fig-xr}, dashed-red and dotted-blue lines show pNFW models based on the best-fit parameters in Table~\ref{tab-ss}.  Lavender polygons represent the best-fit model of Singh et al. (2018). 
	}
	\label{fig-nel_comp}
\end{figure*}

The observation of resolved profiles from individual systems are ideal to constrain $t_{\rm cool} / t_{\rm ff}$. However, such observations suffer from small number statistics due to the small sky coverage of high resolution X-ray/SZ telescopes. These resolved systems are generally biased towards the brightest ones. Additionally, the X-ray emission sharply declines at large radii, making it even more difficult to detect and/or resolve X-ray emission out to large radii and from smaller mass systems. Therefore, the integrated and stacked measurements from lower resolution, large sky surveys such as the ones used in this work represent a practical solution to overcome these issues. In this section, we compare the predictions for the electron-density profiles from our analysis with some of the currently available resolved profiles of individual galaxy cluster, groups and galaxies. We also discuss the biases affecting the individual versus the stacked measurements.

Comparing the electron-density profiles predicted by the analysis of stacked data with direct observations shows that the best fitting models systematically underpredict the observed profiles of massive halos. Figure~\ref{fig-nel_comp} summarizes the situation.  

Each panel of Figure~\ref{fig-nel_comp} shows data representing a different range in halo mass, along with dashed and dotted lines showing model profiles.  Brown lines are the cosmological baseline profiles derived from $K_{\rm base}$ for the stated halo mass.  Pink lines are the precipitation-limited profiles derived from $K_{\rm pre}$ by choosing $t_{\rm cool} / t_{\rm ff} = 10$ and assuming the same metallicity (0.3 $Z_{\odot}$) as the analysis of stacked data. Grey lines give the pNFW profiles derived from adding $K_{\rm base}$ and $K_{\rm pre}$ to obtain $K_{\rm pNFW}$.  In halos with $M_{500} > 10^{14} \, M_\odot$, the pNFW profiles with $t_{\rm cool} / t_{\rm ff} = 10$ are essentially cosmological at $r > 0.1 r_{500}$.  But as halo mass declines, the precipitation limit becomes increasingly restrictive, making the pNFW profiles at the low end of the mass range nearly identical to $K_{\rm pre}(r)$.

Figure~\ref{fig-nel_comp} also shows the models from Section \ref{sec-model}, given the best-fitting parameters determined from our joint fits to stacked X-ray and tSZ data (see Table~\ref{tab-ss}). Dashed-red and dotted-blue lines show the best fitting pNFW models to the stacked data in Figure~\ref{fig-xr}.  Lavender polygons represent the best-fitting model from \cite{singh18} to a subset of the same data spanning a narrower mass range.

\subsubsection{Galaxy Clusters}

In the top left panel of Figure~\ref{fig-nel_comp} are density profiles of galaxy clusters from \cite{cavagnolo09}, which inspired the original precipitation-limited models.  At large radii, those profiles converge to the cosmological baseline profiles. At smaller radii, electron density does not exceed the precipitation-limited profiles with $t_{\rm cool} / t_{\rm ff} = 10$. Cool-core clusters (CC, thin blue lines) tend to track the pNFW models with $\min(t_{\rm cool} / t_{\rm ff}) = 10$.  Clusters without cool cores (NCC, thin red lines) have central densities approximately an order of magnitude less than the cool-core clusters.

Our best-fitting models to the stacked data prefer values of $\min(t_{\rm cool} / t_{\rm ff})$ several times greater than observed in cool-core clusters but less than the values of $\min(t_{\rm cool} / t_{\rm ff})$ observed in some clusters without cool cores.  However, the central density profiles of the NCC clusters are distinctly flatter than in the pNFW models.  As a result, electron density at $r > 0.1 r_{500}$ in essentially all of the galaxy cluster observations exceeds both the best fitting $\gamma$-model and Flat $P$ model.

This feature helps to explain why those models tend to underestimate stacked X-ray luminosity at the high-mass end of our explored range. By construction, a large value of the $t_{\rm cool} / t_{\rm ff}$ parameter suppresses electron density at large radii as well as small radii.   Figure \ref{fig-nel_comp} shows that the properties of hot halo gas in galaxy clusters beyond $0.1 r_{500}$ are determined almost entirely by cosmological structure formation (i.e. $K_{\rm base}$, which by definition is independent of $t_{\rm cool} / t_{\rm ff}$), not the physics of radiative cooling and feedback (i.e.  $K_{\rm pre}$, which carries the dependence on $t_{\rm cool} / t_{\rm ff}$). 
Therefore, the $L_X$  predictions of pNFW models at the high end of our mass range are unrealistically sensitive to the $t_{\rm cool} / t_{\rm ff}$ parameter. For $t_{\rm cool} / t_{\rm ff} >> 10$, they systematically underestimate the electron density observed at $0.1R_{200}$ in galaxy clusters. Consequently, the best-fitting pNFW models to lower-mass systems systematically underpredict $L_X$ for $M_{500} > 10^{14} M_{\odot}$.

\subsubsection{Galaxy Groups}

In the top right panel of Figure~\ref{fig-nel_comp} are density profiles of galaxy groups from \cite{Sun2009}, represented by thin green lines. A pNFW model with $t_{\rm cool} / t_{\rm ff} = 10$ bounded by a $\gamma$-model pressure profile beyond $r_{200}$ traces the upper envelope of the observed profiles, including the inflection in slope near $0.1 r_{500}$, where the profiles go from being precipitation-limited to cosmologically limited.  However, a large majority of the profiles exceed the best-fitting models to the stacked data at small radii, and all of them exceed those best-fitting models at large radii.\\

These systematic differences between direct observations and the fits to stacked data indicate the presence of bias in either the direct observations, the fitting of the stacked data, or both.  The objects analysed by \cite{Sun2009} are 
X-ray selected and therefore are subject to Malmquist bias, meaning that their electron-density profiles are likely to be greater than average for their halo mass.  However, the low scatter among their density profiles at large radii suggests that the true mean density profile nevertheless exceeds the best-fitting model to the stacked data at $> 0.1 r_{500}$.

It is also possible that Eddington bias causes the best fits to the stacked data to have values of $\min(t_{\rm cool} / t_{\rm ff})$ that are larger than the true mean value, therefore underestimating the true electron-density profiles.  Section~\ref{sec-Eddington} discusses this possibility in more detail.

\subsubsection{Early Type Galaxies}

The bottom left panel of Figure~\ref{fig-nel_comp} shows density profiles of early type galaxies from \cite{Babyk2018}. They extend only to $\sim 0.1 r_{500}$ because the low surface brightness of their extended X-ray emission is extremely challenging to observe beyond that radius.  Again, a large majority of the directly observed profiles in this X-ray selected sample exceed the best-fitting models to the stacked data, indicating some form of bias. 

\subsubsection{Milky Way Halo}

The bottom right panel of Figure~\ref{fig-nel_comp} compares the best-fitting models with an array of constraints on the ambient density of the Milky Way's hot halo gas, compiled by \cite{voit19}.  Diamond-shaped polygons represent constraints from ram-pressure stripping of dwarf galaxies.  The brown strip shows constraints from X-ray observations of O~VII and O~VIII absorption lines, and the cyan strip shows constraints from X-ray observations of O~VII and O~VIII emission lines.  For both kinds of X-ray constraints a metallicity of 0.3$Z_{\odot}$ has been assumed. The grey line with inverted triangles marks an upper limit on electron density derived by Anderson \& Bregman (2010) from dispersion-measure observations of pulsars in the Large Magellanic Cloud. As in each of the other panels, the data indicate an electron-density profile with a slope similar to the pNFW model but with a normalization that is greater by a factor $\sim 2$ to 3.

\subsection{Core-excised X-ray luminosity}

In addition to measuring the total X-ray luminosity coming from within $R_{500}$, \cite{anderson15} measured the core-excised\footnote{\cite{anderson15} use the term CGM luminosity instead of core-excised luminosity for both individual galaxies and galaxy clusters. However, we use the latter terminology to avoid confusion as the CGM term is generally associated with only the gaseous medium around individual galaxies.} X-ray luminosity ($\rm L^{ce}_{0.5-2 \, {\rm keV}}$) by masking the central $0.15 \times R_{500}$ region. The core-excised luminosities of galaxy clusters are known to be more robust against the complexities caused by small-scale physical processes compared to the total luminosity and hence are more consistent with self-similar cosmological models \citep{Maughan2007,pratt09, Maughan2012, bulbul18}. 

\begin{figure}
	\includegraphics[height=7.5cm,angle=0.0 ]{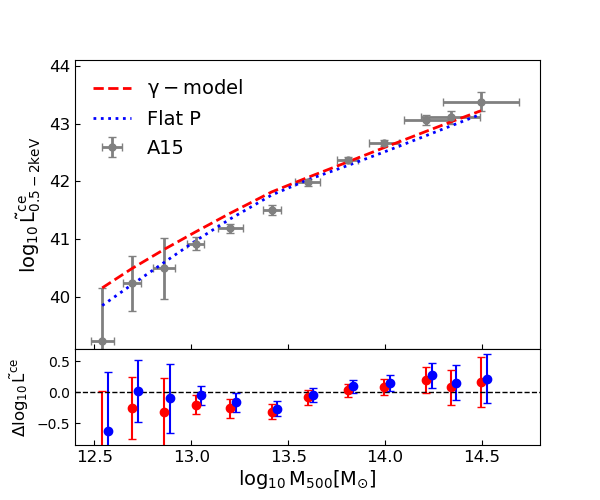}
	\caption{Same as the lower panel of Figure \ref{fig-xr} but for core-excised X-ray luminosity.}
	\label{fig-xr-cgm}
\end{figure}

Figure \ref{fig-xr-cgm} compares the core-excised luminosity measurements by \cite{anderson15} with the predictions of our best-fitting models. We find that both the $\gamma$-model and the Flat~$P$ model fit the data better at the lower mass and the higher mass ends of the data sets, whereas they overpredict the core-excised X-ray luminosity of intermediate mass systems, similar to what we found in case of total X-ray luminosity.

Simulations of X-ray emission from the CGM by \cite{voort16} show that the CGM X-ray luminosity from galaxies with $\ge 10^{12.5}$ M$_\odot$ is independent of star formation rate, and derives mostly from the cooling of hot gas in quasi-hydrostatic equilibrium in the background potential well (see also \citealt{sarkar16}). Therefore the processes operating in the CGM of these galaxies is expected to be similar to those in high mass groups and clusters of galaxies. 

\subsection{Discrepancies near $10^{13.5} \, M_\odot$}
\label{sec-intermediate}
In Figures \ref{fig-xr} and \ref{fig-xr-cgm}, the stacked X-ray luminosities of haloes in the intermediate mass bins ($M_{500} \sim 10^{13.2} - 10^{13.5} M_{\odot}$) lie below the best-fitting model predictions. 
In fact, the dip (i.e. the residual) in the core-excised luminosity is more pronounced than the dip in total luminosity ($2.5\sigma$ in the case of core-excised X-ray luminosity as compared to $2\sigma$ in the case of total X-ray luminosity) at these intermediate masses indicating that the processes responsible for the discrepancy play an important role in gas thermodynamics even at $r > 0.15\times R_{500}$.
This dip in the X-ray luminosity may be related to the high central velocity dispersion ($\sigma_v \gtrsim 240 \rm \, km\, s^{-1}$) found in massive ellipticals, which in turn maximizes the efficiency of AGN feedback \citep{Voit2020}. In these systems, the AGN power is sufficient to reduce the CGM pressure by lifting hot halo gas out of the galaxy's potential well, thus reducing X-ray luminosity. This effect cannot persist to massive groups and galaxy clusters, because the greater pressure of the surrounding medium does not allow as much lifting of the halo gas by feedback. And at lower masses, kinetic AGN power is not captured as effectively by the CGM and tends to drive multiphase circulation instead of lifting the CGM out of the galactic potential well.

This effect is also highlighted in Figure 5 of \citet{Voit2020} in which the entropy profiles of massive elliptical galaxies with $\sigma_v \gtrsim 240 \, {\rm km \, s^{-1}}$ are substantially greater than the precipitation limit at $\sim$ 10 kpc, translating to a lower CGM density. 
In the observational study of X-ray bright elliptical galaxies, \cite{Lakhchaura2018} found cool gas free (i.e. single phase) ellipticals have higher $t_{\rm cool}/t_{\rm ff}$ ($\approx 40-90$) in the radial range $\sim$ 3-35 kpc than the ratio ($\approx 30$) in ellipticals with extended multiphase gas. \cite{Voit2020} showed that the absence of multiphase gas is correlated with a greater velocity dispersion, and therefore these are the same type of galaxies with higher values of $t_{\rm cool}/t_{\rm ff}$.
Our pNFW model therefore does not apply to these systems and forcing the fiducial model to go through those data-points biases the results. 

We therefore repeated the MCMC analysis excluding the two mass bins centered at $M_{500} = 10^{13.20}$ and $10^{13.42} M_{\odot}$. Table \ref{tab-ss} shows the results of that analysis. There are only negligible changes in $f_T$ and $\alpha_T$ and hence the temperature profiles. However, the best-fitting value of $t_{\rm cool}/t_{\rm ff}$ decreases by $\sim 10\%$ (for both the models presented in this paper), consistent with a small rise in amplitude of the $L_{0.5-2 \, {\rm keV}}-M_{500c}$ scaling relation that would otherwise be lowered by the inclusion of intermediate mass halos.

\subsection{Eddington bias}
\label{sec-Eddington}
The presence of scatter in the observable-mass scaling relations (in our case $L_{0.5-2 \, {\rm keV}}-M_{500c}$ and $M_*-M_{500c}$) has an indirect impact on the estimation of stacked luminosity which can potentially bias our estimates of best-fitting model parameters. The net change in inferred X-ray luminosity (in a given stellar mass bin) compared to an ideal model depends on a number of observationally well constrained ones (e.g., the slope of the halo mass function) as well as some poorly constrained quantities (e.g., a potential anticorrelation between $L_{0.5-2 \, {\rm keV}}$ and $M_*$ at fixed halo mass).  To first order, it is given by,
\begin{equation}
\begin{split}
\label{eqn-dellx2}
  \Delta \langle \log L_{0.5-2 \, {\rm keV}} \rangle = \log e 
    \times [- \alpha_{lx}\beta \sigma^2 _{\mu|m_*} 
    \; \; \; \; \; \; \\ \; \; \; \; \; \;  
    + \: \alpha_{lx}\beta r_{lx,m*} \sigma_{\mu|l_x} \sigma_{\mu|m_*}]
\end{split}
\end{equation}
where, $\beta$ is the local slope of halo mass function, $r_{lx,m*}$ is the covariance between $L_{0.5-2 \, {\rm keV}}$ and $M_*$, and $\sigma^2 _{\mu} \equiv \sigma^2 / \alpha^2$, with the $\alpha$s and $\sigma$s being the mass slopes and log-normal scatter of the scaling relations, respectively. Appendix \ref{apn-dellx} provides a
detailed derivation of Equation \ref{eqn-dellx2}, based on \cite{Evrard_2014MNRAS.441.3562E}.

The first term in Equation \ref{eqn-dellx2} represents Eddington bias. Overall, there are more low mass (and low $L_X$) halos, and therefore a greater number are scattered into higher stellar-mass bins, compared to the number of high mass (and low $L_X$) halos that scatter into the lower 
stellar-mass bins. That mismatch reduces the mean X-ray luminosity attributed to the halo mass associated with a given stellar-mass bin. The second term is controlled by the correlation between X-ray luminosity and stellar mass at fixed halo mass \citep[e.g.][]{Farahi_2019NatCo..10.2504F}. A positive correlation ($r_{lx,m*}>0$) enhances the mean X-ray luminosity, 
whereas, a negative correlation ($r_{lx,m*} < 0$) reduces it. Also, $r_{lx,m*}$ is the quantity least well constrained by observations in Equation \ref{eqn-dellx2}.

The value of $\beta$ depends on the mean mass and redshift of the halo mass bin and it varies between $\sim 2-3$ from the lowest to the highest mass bin considered in this paper. We use $\alpha_{lx}=1.93$ and $\sigma_{lx} = 0.25$ from \cite{bulbul18}, and $\alpha_{m*}=0.8$ and $\sigma_{m*} = 0.22$ from \cite{chiu2018}. Note that both of these studies involve massive galaxy clusters ($M_{500} > 2-3 \times 10^{14} M_{\odot}$) spanning a wide redshift range. We fix $r_{lx,m*} = -1$, and the slope and scatter of the scaling relations to these ballpark values to get a qualitative estimate of maximum possible decrement in the mean X-ray luminosity. We find $\Delta \langle \log L_{0.5-2 \, {\rm keV}} \rangle \sim -0.2$ at $M_{500} < 10^{13.8} M_{\odot}$. The decrement quickly increases for higher mass halos reaching $\Delta \langle \log L_{0.5-2 \, {\rm keV}} \rangle \sim -0.3$ at $M_{500} = 10^{14.5} M_{\odot}$. 

Repeating the MCMC analysis after taking into account the above decrements in X-ray luminosity gives $t_{\rm cool}/t_{\rm ff} = 71.44^{+6.85}_{-6.49}$, $f_T = 0.86 \pm 0.06$ and $\alpha_T = -0.13 \pm 0.04$ i.e. approximately 30\% and 10\% decrements in first two parameters, respectively, and a negligible change in $\alpha_T$. This experiment highlights that the large value of $t_{\rm cool}/t_{\rm ff}$ can partly be explained by the presence of Eddington bias in the stacked data sets. However, it is insufficient to bring our predictions for the stacked measurements in agreement with the ones obtained for X-ray selected sample of individual groups and clusters.\\

\section{Summary}
\label{sec-summary}

This paper has presented a simple analytical model for the hot diffuse halo gas, inspired by the precipitation limit that appears to place an upper bound on the gas-density profiles of galaxy clusters and groups. We tuned the model parameters to the fit stacks of integrated X-ray and tSZ data from halos spanning two orders of magnitude in mass ($M_{500} \sim 10^{12.5}-10^{14.5} M_{\odot}$) to see if this single halo-gas model could unify the ICM and CGM. Here, we summarize our main findings. 
\begin{enumerate}
    \item Fitting of our analytical model with a $\gamma$-model boundary condition to the stacked data gives the constraints $t_{\rm cool}/t_{\rm ff} = 104.06^{+9.74}_{-9.40}$, $f_T = 0.95\pm0.06$ and $\alpha_T = -0.14 \pm 0.03$. The negative value of $\alpha_T$ means that the ratio of gas temperature at $R_{200}$ to the virial temperature of the halo decreases as the halo mass increases.
    \item For the analytical model with a Flat $P$ boundary condition, the best-fit model parameters drop to $t_{\rm cool}/t_{\rm ff}= 63.23^{+4.43}_{-4.34}$ and $f_T = 0.36 \pm 0.02$, while $\alpha_T$ remains unchanged. 
    \item  These statistically significant differences in the best-fit parameters indicate that the systematic uncertainites in our modeling exceed the statistical ones. Uncertain assumptions about the pressure profile beyond the virial radius are necessary to calculate a cylindrical tSZ signal within the $5 R_{500}$ aperture of the {\it Planck} data. Those assumptions affect our best-fitting parameters through their influence on the temperature boundary condition at $R_{200}$, leading to a degeneracy between the $t_{\rm cool}/t_{\rm ff}$ and $f_T$ parameters.    
    \item The observed hot-gas density profiles in individual galaxy clusters, galaxy groups, massive ellipticals, and maybe even the Milky Way are typically a factor $\sim 2$--3 greater than the predictions of our best-fitting models to the stacked data. This discrepancy could arise from two different biases, one affecting the selection of the directly observed objects (e.g., Malmquist bias), and another that arises from the steep mass distribution function of the stacked data sets we use (e.g., Eddington bias). However, it is clear that the pNFW model with $t_{\rm cool}/t_{\rm ff}=10$ traces the upper envelope of the directly observed density profiles across a wide range of halo masses.
    \item Our rough assessment of the magnitude of the Eddington bias indicates that accounting for the maximum possible amount of that bias reduces the best-fitting $t_{\rm cool}/t_{\rm ff}$ parameter by as much as by 30\% and reduces $f_T$ by 10\%.
\end{enumerate}

Our analysis demonstrates that the pNFW model can match to the stacked data-sets at both the high-mass end (i.e. massive galaxy groups and galaxy clusters) and the low-mass end (i.e. individual massive galaxies),
thus providing a unified prescription for the CGM/ICM. However, it overpredicts the X-ray luminosity at intermediate masses ($M_{500} \sim 10^{13.5} M_{\odot}$), indicating that the simplistic nature of the model is unable to capture the physical processes shaping the gas distribution at these mass scales. 
We expect the latest tSZ and kSZ measurements at intermediate halo masses (\citealt{Schaan2020}) to play a crucial role in understanding the complex interplay between AGN feedback and the surrounding gas for such systems. Besides, cosmological hydrodynamical simulations such as Magneticum \citep{dolag2015,Teklu2015} exploring individual galaxies, groups and galaxy clusters with simulation boxes varying in size and resolution (eg. from medium resolution boxes exploring $\sim$Gpc scales to extremely high resolution boxes at smaller scales from tens of Mpc to tens of kpc) are expected to help in bridging the gap in the unification of the gas physics from the CGM to the ICM.\\\\\\
{\bf{ACKNOWLEDGEMENTS}}\\
	We thank the anonymous referee for valuable suggestions and comments. We thank Wenting Wang for generously sharing the effective halo mass estimates from their stacked weak-lensing measurements. We thank Daisuke  Nagai, Alex Saro and Prateek Shrama for their valuable suggestions.
	PS is supported by the ERC-StG ‘ClustersXCosmo’ grant agreement 716762.\\\\\
{\bf{DATA AVAILABILITY}}\\
	The data-sets underlying this article are either publicly available or via the authors of their respective publications. 
	\footnotesize{
		\bibliography{bibtexcgm}{}
		\bibliographystyle{mn2e}
	}

\begin{appendices}
\renewcommand\thefigure{\thesection.\arabic{figure}}    
\renewcommand{\thetable}{A\arabic{table}}
\renewcommand{\theequation}{A\arabic{equation}}

\section{Eddington bias}
\setcounter{figure}{0}    
\setcounter{table}{0}
\setcounter{equation}{0}
\label{apn-dellx}
To estimate Eddington bias, we follow \cite{Evrard_2014MNRAS.441.3562E} by assuming power-law scaling relations for $L_{0.5-2 \, {\rm keV}}-M_{500c}$ and $M_*-M_{500c}$ expressed in logarithmic form,
\begin{equation}
\langle l_X| \mu \rangle  = \pi_{lx} + \alpha_{lx} \mu \pm \sigma_{lx} \nonumber
\end{equation}
\begin{equation}
\langle m_*| \mu \rangle = \pi_{m*} + \alpha_{m*} \mu \pm \sigma_{m*}
\end{equation}
where, $\ln L_{0.5-2 \, {\rm keV}} \equiv l_X$, $\ln M_* \equiv m_*$, $\ln M_{500} \equiv m_{500}$, $\mu \equiv \ln \frac{M_{500}}{M_p}$, 
$M_p$ is the pivot mass, $\pi$, $\alpha$ and $\sigma$ are the normalization, mass slopes and 
log-normal scatter of the scaling relations, respectively. 
The first order expansion of the mass function around pivot halo mass is given by,
\begin{equation}
n_1 (\mu) = A e^{-\beta \mu}
\end{equation}
where, $A$ and $\beta$ are local amplitude and slope of the mass function, respectively. Using above approximation to obtain the mean X-ray luminosity at fixed stellar mass gives,
\begin{equation}
\label{eqn-lxms}
\langle l_X|m_*\rangle = \pi_{lx} + \alpha_{lx} \Bigl[ \langle \mu |m_* \rangle + \beta r_{lx,m*} \sigma_{\mu|l_x} \sigma_{\mu|m_*} \Bigr]
\end{equation}
where, $r_{lx,m*}$ is the covariance between $L_{0.5-2 \, {\rm keV}}$ and $M_*$, $\sigma^2 _{\mu|l_x,1} = \sigma^2 _{lx}/ \alpha^2 _{lx}$, $\sigma^2 _{\mu|m_*} = \sigma^2 _{m_*}/ \alpha^2 _{m_*}$, and 
\begin{equation}
\label{eqn-mums}
\langle \mu |m_*\rangle = (m_* - \pi_*)/\alpha_{m_*} - \beta \sigma^2 _{\mu|m_*}
\end{equation}
Combining equation \ref{eqn-mums} and \ref{eqn-lxms} gives the mean X-ray luminosity in a given stellar mass bin i.e., 
\begin{equation}
\label{eqn-lxms2}
\langle l_X|m_*\rangle = \pi_{lx} + \alpha_{lx} \Bigl[(m_* - \pi_*)/\alpha_{m_*} - \beta \sigma^2 _{\mu|m_*} + \beta r_{lx,m*} \sigma_{\mu|l_x} \sigma_{\mu|m_*} \Bigr]
\end{equation}
In the absence of any scatter in the scaling relations we obtain,
\begin{equation}
\label{eqn-lxms3}
\langle l_X|m_*\rangle = \pi_{lx} + \alpha_{lx} (m_* - \pi_*)/\alpha_{m_*}
\end{equation}
Therefore, the presence of scatter changes the mean luminosity in a given stellar mass bin by,
\begin{equation}
\label{eqn-dellx}
\Delta \langle l_X|m_*\rangle = - \alpha_{lx}\beta \sigma^2 _{\mu|m_*} + \alpha_{lx}\beta r_{lx,m*} \sigma_{\mu|l_x} \sigma_{\mu|m_*}
\end{equation}

\end{appendices}
	
\end{document}